\documentclass[aps,showpacs,twocolumn]{revtex4}

\usepackage{bm}
\usepackage{amsmath}
\usepackage{amssymb}
\usepackage{graphicx}
\usepackage{epsfig}
\usepackage{epstopdf}
\begin{document}

\def\be{\begin{equation}}
\def\en#1{\label{#1}\end{equation}}
\def\d{\dagger}
\def\bar#1{\overline #1}
\newcommand{\per}{\mathrm{per}}

\newcommand{\rd}{\mathrm{d}}
\newcommand{\vare}{\varepsilon }

\title{ Conditions on the experimental  Boson-Sampling computer to disprove   the Extended Church-Turing thesis  }

\author{V. S. Shchesnovich}

\address{Centro de Ci\^encias Naturais e Humanas, Universidade Federal do
ABC, Santo Andr\'e,  SP, 09210-170 Brazil }

\begin{abstract} 
We give a set of sufficient conditions on the experimental Boson-Sampling computer   to satisfy  Theorem 1.3  of Aaronson $\&$  Arkhipov (Theory of Computing \textbf{9}, 143 (2013))  stating  a computational problem whose simulation on a classical computer would collapse the polynomial  hierarchy of the computational complexity to the third level. This implies that such an experimental  device is  in conflict with the Extended Church-Turing thesis.       In practical terms,  we give a set of  sufficient conditions for the  scalability  of the experimental    Boson-Sampling computer    beyond the power of  the classical computers. The derived  conditions can be also used for devising efficient verification tests of the Boson-Sampling computer. 
\end{abstract}

\maketitle

\textit{Introduction --} The Boson-Sampling (BS) computer  was recently proposed  by  S. Aaronson and A. Arkhipov \cite{AA} as a near-future feasible device serving as an evidence  against the Extended Church-Turing thesis (ECT), i.e., that any physical device can be \textit{efficiently} simulated  on a classical computer.  The physical setup of  the BS computer in the  linear optics realization   is a significantly reduced version   of the  Knill, Laflamme \&  Milburn (KLM)  universal quantum computer (UQC)   \cite{KLM}.    The optical BS  device consists of an  unitary  linear network,  with the     indistinguishable  single photons    producing  the Hong-Ou-Mandel type interference \cite{HOM} (see also Refs. \cite{LB,MPI})   at the network input, and non-adaptive photon counting measurements at the network output.  Ref. \cite{AA} shows evidence that \textit{even an approximate} classical simulation of the probability distribution in the   BS computer output would  collapse the hierarchy of the computational complexity to the third level. The latter is an implausible consequence,  therefore  the BS computer  is in conflict  the ECT.  

An UQC could  simulate the BS computer, but the scalability of the BS device  beyond the classical computational power could be  easier to achieve. Indeed,  the sources of the single photons make spectacular advances \cite{SPS} and only the passive optical elements are needed. In practical terms, with   few dozens of single  photons the BS device would outperform  the current classical computers \cite{AA}.       Hence, though it is  not known if any practical computational task can be solved on the BS computer, such a device  undoubtedly would have an enormous   impact on physics.  Four independent groups have   tested their prototypes  on small networks with few single  photons    \cite{E1,E2,E3,E4}. Now the goal is to    scale up the BS computer  to at least few dozens of single photons.

The most important novelty associated with    Aaronson $\&$  Arkhipov's proposal of the BS computer is \textit{a change  of the focus}: Instead of asking a quantum device to efficiently solve an  arithmetic  problem, such  as factoring large numbers \cite{Shor,BookNC}, Aaronson $\&$  Arkhipov  propose to  look for   the realistic   quantum devices with   \textit{the  output itself} being a classically hard computational problem.  They then argue that the  BS computer is one of such quantum systems. The main reason is that in  the ideal case (with perfectly indistinguishable single photons, noiseless unitary network and ideal detectors)    the  $N$-boson   output  amplitudes   are  given as the  matrix permanents  (see Ref. \cite{Minc}) of complex $N\times N$-submatrices of the network matrix \cite{C,S} and,  by the  classic result of the computation complexity theory  \cite{Valiant},   require exponential in $N$ computation time  (see also Ref. \cite{A1}). The fastest known  algorithm  for computation of the matrix permanent, due to H. Ryser \cite{Ryser}, requires $O(N 2^N)$ operations.         The   matrix permanent is a classically hard problem in a   superior computational complexity class, the class $\#$P (i.e., problems, generally,  with no efficient classical  algorithm  even for approximation of the solution).   Using this, Aaronson $\&$  Arkhipov \cite{AA} present arguments   that the  complexity of the  BS computer operation  is  \textit{stable}  under errors. They argue  that  simulation of  the probability distribution in the BS computer output  is classically hard even with  the variational distance error being    \textit{an integral part}  of the problem.  This is a novel feature:  previously non-universal quantum  computations  \cite{TV,SB,BJS} concerned  the   \textit{idealized} quantum systems. 

The stability of the classical hardness of the BS computer output under errors    is  the key property. Indeed, the UQC aims at  problems in the NP class of the computational complexity  (i.e., with the solution being easily verifiable on a classical computer) and requires  the error correction. The discovery  of the error correction protocols \cite{ErrCorr1,ErrCorr2} was a decisive step in favor of the physical feasibility  of the UQC.   Whereas, the probability distribution in a non-zero  variational distance to that of the ideal BS computer  output is \textit{itself} a classically hard problem. Such  a variational distance    enters the description of a  classically hard problem in Theorem 1.3 of  Ref.  \cite{AA}, the main result of  Aaronson $\&$  Arkhipov.    If   it were  not the case, the BS computer  would  be quite similar to  the  analog classical computer which is NP-powerful in the ideal case, but  in reality the  errors in the physical setup   reduce its power to that of the usual (binary) classical computers.

Thus, the BS computer is a quite different type of computer, to which the standard notions of an error and  error correction do not apply.  Hence,   application of the techniques from the realm of the UQC is not straightforward. For instance,  the  conclusion made in Ref. \cite{ConstErr} that the BS computer cannot disprove the ECT is ungrounded.  First,  the  errors   in the experimental setup were postulated  $N$-independent (where $N$ is the number of the photon sources), which is  an unjustified  restriction for such a general claim. It was known before  that the scalability requires $1/\textrm{poly}(N)$ scaling of   errors in the experimental setup, see Refs.  \cite{NoisyBS, NDBS}.  Second,  the authors of Ref. \cite{ConstErr}  completely  ignore  Theorem 1.3 of Ref. \cite{AA} by implicitly postulating in their premises  that   the BS computer  output  is classically hard only in  the exact runs, corresponding to the ideal BS computer~\footnote{The conclusion of Ref. \cite{ConstErr} is a byproduct of  two unjustified  restrictive  assumptions: (i) that      only the exact runs, i.e.,  corresponding to the ideal BS computer,  correspond to  a classically hard problem, (ii) that, moreover, the setup parameters are kept $N$-independent.}.
 
 \textit{Theorem 1.3 of Ref. \cite{AA} --}  Adopted to our case, it states: Let $\mathcal{D}$ be the probability distribution sampled by the ideal BS computer $\mathcal{I}$.   Suppose that there exists a classical algorithm $C$ that takes as input the  description of $\mathcal{I}$ as well as an error bound $\epsilon$  on the variational distance $||\ldots||$~\footnote{The variational distance between two probability distributions, say $p_i$ ($\mathcal{P}$) and $q_i$ ($\mathcal{Q}$), is defined as  $||\mathcal{P}-\mathcal{Q}||\equiv \sum_{i}|p_i-q_i|$.} and  samples from a probability distribution $\mathcal{D}^\prime$ such that $ ||\mathcal{D}^\prime-\mathcal{D}||\le \epsilon$ in $\mathrm{poly}(|\mathcal{I}|, 1/\epsilon)$ time. Then the absolute values of  permanents of complex Gaussian matrices can be efficiently approximated on a classical computer.   This, under two highly plausible numerically tested  conjectures,  implies  collapse of the polynomial hierarchy to the third level.

Theorem 1.3  refers to the ideal BS computer in the so-called ``collision free" regime, i.e.,  with  $N$ indistinguishable single  photons on a network with $M$ modes such that $M\gg N^2$.   Only one of the two conjectures (the Permanent-of-Gaussians  conjecture and the Permanent Anti-Concentration conjecture) has to do with the computational complexity theory (the Permanent-of-Gaussians)~\footnote{Calculation of  the permanent of matrices of  mutually independent complex  Gaussian  random elements       to a multiplicative error is a classically hard problem.  Similarly,  the classical  hardness of factoring of large numbers is also only a conjecture. }.

\textit{ The  experimental BS computer  and  the ECT --}  The conceptual importance of the  BS computer  for physics lies in its    obvious conflict with the  ECT   \cite{AA}.  Indeed,   by Theorem 1.3,  given a random $M$-mode  network with  $N$  of  its  input modes   connected to the  ideal single photon sources,  it is   impossible  to efficiently simulate the output distribution of this  ideal  BS computer on a classical computer  (in poly$(M,N,1/\epsilon)$ time, where  $\epsilon$  is a given variational distance error). Therefore,  if we find the conditions on the experimental setup of a realistic BS computer, which would guarantee that   the   computational  problem of Theorem 1.3 is \textit{efficiently} simulated  by such an experimental  BS device, then this would be the first step to falsify the ECT (building an operating   BS device  is the second, decisive, step).    

Below   a  set of sufficient conditions is  given for the experimental   BS  device to simulate the output of the ideal BS computer to a  variational distance  error  $\epsilon$    for the fraction $1-\delta$ of all networks, for any $N$, $M$, and  $\epsilon,\delta>0$, i.e., we give the  conditions for   scalability  of the experimental BS computer, as required for falsification of the ECT, where one has to use  $1/(1-\delta)$ different networks, on average  to have a hard instance of the network. The simulation is efficient since   no postselection is used  on the experimental BS computer runs.

Here we note that    necessary  conditions  for the BS  computer to operate beyond the classical  computation  power were discussed   in Refs. \cite{RR,R1}.    Sufficient  condition on  noise in the  optical network  was found in Ref.  \cite{NoisyBS}:  the  fidelity of  the optical elements in a noisy network  must be  $\mathcal{F}_{el} = 1- O(N^{-2})$.   In Ref. \cite{NDBS}  it was shown that   the BS computer  is scalable  when  the average  single-photon  fidelity between any pair of the  photon sources satisfies  $\langle \mathcal{F}_{ph}\rangle  = 1- O(N^{-3/2})$.   

In the limit of small setup errors one can consider  them separately.  Hence,  by  neglecting the effect of the photon mode mismatch,  we   consider  below    the combined effect of  multi-photon components in the input modes,  photon losses,   and  detector  dark counts. Then, we  return to   the effect of the mode mismatch, neglecting  the above  errors.

\textit{Effect of  multi-photon components, photon losses, and detector dark counts --}  Assume for a while that the network is  ideally  unitary and the  photons are ideally indistinguishable (i.e., have  the same spectral function \footnote{We neglect polarization of the photons, assuming it to be fixed.}).      Let  $N$ photon sources,  replicas of each other, each output the density matrix    $\rho^{(i)} = p_0 |0\rangle\langle0| + p_1\rho^{(i)}_1 + p_2\rho^{(i)}_2 + ... $, where index $i$ labels the spatial mode of the network  and the $k$-photon component  $\rho^{(i)}_k$  has the probability $p_k$.   The  input   density matrix (with  the input modes  $1,\ldots, N$ connected to the photon sources)  reads 
\be
\rho = \rho^{(1)}\otimes\ldots\otimes\rho^{(N)}\otimes|0\rangle\langle0|\otimes\ldots\otimes|0\rangle\langle0|. 
\en{E1}
Set   $a_i$ and $b_i$,  $i=1,\ldots,M$ to be  the boson operators for the input and the output modes of the network. We have   $a^\dag_i = \sum_{l=1}^M U_{il}b^\dag_l$, with an unitary  matrix $U$.  Since for now we neglect the mode mismatch,   $\rho^{(i)} \equiv \sum_{k=0}^\infty {p_{k}}/{k!}(a^\dag_i)^{k}|0\rangle\langle0| a^{k}_i$.  Photon  losses can be accounted for by introduction of the   loss probability $r$ (and considering them to occur at the detection stage), whereas the dark counts can be  described by the integral  dark count  rate   $\nu$  \cite{Barnett,Lee}.    $M$ bucket detectors, replicas of each other,  connected to the network output   are described   by the no-click probability   $P_D(0|s) = e^{-\nu}r^s$, for the  $s$-photon input (i.e., the zero dark counts probability $e^{-\nu}$ multiplied by the total loss probability $r^s$), and the click probability $P_D(1|s) = 1 - e^{-\nu}r^s$.

It is  convenient to introduce a vector notation for the mode occupation numbers,  writing $|\vec{n}\rangle$ for  the  Fock state with $\vec{n} =(n_1, \ldots,n_M)$. Let us set  $|\vec{n}| = \sum_{i=1}^M n_i$. In Eq. (\ref{E1}) we have the input  $\vec{n}$ with  $n_{i} \ge 0$, for $1\le i \le N$,  and $n_{i}=0$, for  $i\ge N+1$.   The input Fock state $|\vec{n},in\rangle$  expansion in output Fock states $|\vec{s},out\rangle$  reads \cite{A1,AA,C,S}
\be
|\vec{n},in\rangle = \sum_{\vec{s}}\delta_{|\vec{n}|,|\vec{s}\,|} \frac{\per(U[\vec{n}\,|\vec{s}\,])}{\sqrt{\mu(\vec{n})\mu(\vec{s}\,)}}|\vec{s},out\rangle,
\en{E2}
where $\mu(\vec{n}) = \prod_{i=1}^M n_i!$, $\per(\ldots) $ stands for the matrix permanent \cite{Minc}, and  we denote by $U[\vec{n}\,|\vec{s}\,]$ the $N\times N$-dimensional matrix obtained from  $U$ by taking  the $k$th row $n_k$ times and the $l$th column $s_l$ times (the order of rows/columns being unimportant). 

The probability of  $N_o$ clicks of the output detectors located at   $\vec{l}=(l_1,\ldots,l_{N_o})$    reads
\be
P_{out}(\vec{m}\,) =\sum_{\vec{s}}P_D(\vec{m}\,|\vec{s}\,) \sum_{\vec{n}}P_U(\vec{s}\,|\vec{n}\,)P_I(\vec{n}\,),
\en{E3}
where   the  binary ``occupation number" $m_l$ counts the $l$th detector clicks ($m_{l_\alpha} = 1$ for $1\le \alpha \le N_o$ and $m_{l_\alpha} = 0$ for $N_o+1 \le \alpha\le M$).  Here the probability $P_I(\vec{n}\,)$ of the input  $\vec{n}$, the  conditional    probability of the network  output $\vec{s}$, $ P_U(\vec{s}\,|\vec{n}\,)$, and the conditional detection probability  $P_D(\vec{m}\,|\vec{s}\,)$ are  given as follows:
\begin{eqnarray}
\label{E4}
&&   P_U(\vec{s}\,|\vec{n}\,) = |\langle \vec{s},out|\vec{n},in\rangle|^2 = \frac{|\per(U[\vec{n}|\vec{s}])|^2}{{\mu(\vec{n})\mu(\vec{s}\,)}}
\delta_{|\vec{n}|,|\vec{s}\,|},\nonumber\\
&& P_I(\vec{n}\,) =  \prod_{i=1}^Np_{n_i}, \quad P_D(\vec{m}\,|\vec{s}\,)  =  \prod_{l=1}^M P_D(m_l|s_l), \nonumber\\
&& P_D(m|s) = e^{-\nu}r^s\delta_{m,0} + (1-e^{-\nu}r^s)\delta_{m,1}.
\end{eqnarray}
Note the obvious identities: $\sum_{\vec{n}} P_I(\vec{n}\,) = 1$, $\sum_{\vec{s}}P_U(\vec{s}\,|\vec{n}\,)  = 1$, and $ \sum_{\vec{m}}P_D(\vec{m}\,|\vec{s}\,) =1$.

In the   ``collision  free" regime,  $M\gg N^2$, due to the boson birthday paradox \cite{AA, AG, BBPExp},   the probability  of photon bunching at the   network output is bounded, on average in the Haar measure,  by     $1 - (\sum_{|\vec{m}|=N}1)/(\sum_{|\vec{n}|=N}1) =  1- \prod_{k=1}^{N-1}(1-{k}/{M})< N(N-1)/2M$. Thus,  simple   bucket detectors, registering only the presence of   an  input different from the vacuum, are sufficient. 

Denote by $ \mathcal{V}$ the variational distance between the output probability distributions of the realistic and the  ideal BS computers.   As above discussed, the bunched output of  the ideal BS computer contributes the term $\langle \mathcal{V}_b\rangle \le N^2/2M$ (here and below $\langle...\rangle$ means the averaging in the Haar measure). The rest of $\mathcal{V}$ consists of the following two parts:
\be   
\mathcal{V}_1 \equiv  \sum_{|\vec{m}|\ne N} P_{out}(\vec{m}\,), \;
 \mathcal{V}_2 \equiv \sum_{|\vec{m}|= N} |P_{out}(\vec{m}\,) - P_{out}^{(0)}(\vec{m}\,)|, 
 \en{E5}
where $P_{out}^{(0)}(\vec{m}\,) \equiv P_U(\vec{m}\,|\vec{n}^{(0)})$ with  the ideal  input  $\vec{n}^{(0)}$, i.e., $n^{(0)}_i=1$ for $i=1,\ldots,N$ and $n^{(0)}_i=0$ otherwise.    

First of all, in $\mathcal{V}_{1,2}$ of Eq. (\ref{E5}) we have an exponential number of terms in the summation over $\vec{m}$ and also   over $\vec{s}$   and $\vec{n}$  in $P_{out}(\vec{m}\,)$, see Eq. (\ref{E3}). Hence an exponentially small  bound  on the  output probabilities $P_U(\vec{s}\,|\vec{n}\,)$ is needed to bound such a sum.  This can be achieved by employing     Chebyshev's inequality with respect to the Haar measure $Pr(...)$: we simply exclude  a  fraction $\delta$ of all networks.    Chebyshev's inequality reads $Pr(\mathcal{V}  <  \epsilon) \ge  1 -  {\langle \mathcal{V} \rangle}/{\epsilon}$, it  supplies a sufficient condition that  an experimental BS device   is $\epsilon$-close in the variational distance to the ideal BS computer at least for the fraction $1-\delta$   of the network matrices with   $ \delta= \langle \mathcal{V}\rangle/\epsilon$.   We obtain   (see the derivation in  Ref. \cite{Suppl}), recalling also the bunching term,
\be
\langle \mathcal{V}_1+\mathcal{V}_2 +\mathcal{V}_b \rangle \le  \mathcal{R}_A, 
\en{E15A}
where
\be
 \mathcal{R}_A =   \frac{N^2}{2M}+ 2\left\{1-Q\left[1 - \frac{N^2}{2M}\right] \right\} + 1-Q + Q^\prime
\en{RA}
with $Q \equiv e^{-(M-N)\nu}\left(1 - e^{-\nu}r\right)^N p_1^N$ and $Q^\prime \equiv 1- p_1^N$.  Eqs. (\ref{E15A})-(\ref{RA}) have a very clear physical meaning, since $Q$ is the   probability  that  $N$ detectors have clicked,  $M-N$ detectors had  zero  dark counts, and that $N$ indistinguishable single photons were at the  network input, whereas $Q^\prime$ is the   probability of a non-ideal input. Applying Chebyshev's inequality, we get the sufficient condition on the multi-photon components, photon losses, and detector dark counts as  
\be
  \mathcal{R}_A (M,N,p_1,\nu,r)\le \epsilon\delta.
\en{E16}
Finally, we have  $\mathcal{R}_A \le  {N^2}/{M} + 3\left[(M-N)\nu +N r \right]+ 4N(1-p_1)$ \cite{Suppl}, and  a simpler   condition   (sufficient for Eq. (\ref{E16}))   follows 
\be
 \frac{3N^2}{2M} +  3\left[(M-N)\nu +N r \right]+ 4N(1-p_1) \le \epsilon\delta.
\en{E17}
  Eq. (\ref{E17}) reveals that  multi-photon components (and the vacuum), photon  losses, and detectors dark counts are  additive errors  with respect to the classically hard problem size, i.e., $M$ and $N$  (it is not so in the case of the photon mode mismatch).  

\textit{Effect of the  photon mode mismatch  --}  For a realistic  BS computer with only partially indistinguishable photons,    the network  output probability,   generalizing   $P_U(\vec{s}\,|\vec{n}\,)$  of    Eq. (\ref{E4})  for  nonzero mode mismatch and the approach of Ref.~\cite{SU3} for $N>3$,  reads   \cite{Suppl}
\be
\widetilde{P}_U(\vec{s}\,|\vec{n}\,) = \frac{1}{\mu(\vec{s}\,) \mu(\vec{n}\,)}\mathrm{Tr} \left\{ \mathcal{U}\rho^{(1)}\otimes\ldots\otimes \rho^{(N)}  \mathcal{U}^\dag\right\},
\en{P}
where $|\vec{n}|= |\vec{s}\,|$,   Tr($\ldots$) is the trace in the  tensor product of $|\vec{n}|$   Hilbert spaces  associated with the frequencies of  spectral decomposition of the multi-photon states, i.e.,  with the basis $  |\omega_1,...,\omega_{|\vec{n}|}\rangle \equiv |\omega_1\rangle\otimes...\otimes|\omega_{|\vec{n}|}\rangle$,     $\mathcal{U} = \sum_\sigma\left[\prod_{\alpha=1}^NU_{k_{\sigma(\alpha)},l_\alpha}\right]\mathcal{P}^\dag_\sigma$ with the sum running over all permutations $\sigma$ and   the operator $\mathcal{P}_\sigma$ acting  as follows $\mathcal{P}_\sigma |\omega_1,...,\omega_{|\vec{n}|}\rangle \equiv |\omega_{\sigma^{-1}(1)},...,\omega_{\sigma^{-1}(|\vec{n}|)}\rangle$ ($P_\sigma$  permutes    the frequencies in the spectral  expansion   of the multi-photon   states with respect to the spatial indices of the input modes).  Here the input modes $(k_1,\ldots,k_{|\vec{n}|})$ and the output modes $(l_1,\ldots,l_{|\vec{n}|})$ correspond to  the occupation numbers $\vec{n}$ and $\vec{s}$, respectfully. 

We  consider the case of  mixed  $\rho^{(i)}$ and, as above discussed,   neglect  multi-photon  (and vacuum) components,  photon losses, and detector dark counts. Thus   $\rho^{(i)} = \rho^{(i)}_1\equiv \rho_1$ (identical sources) and $|\vec{n}|=N$.  Due to the identical single photon sources, the     permutations $\sigma$ in Eq. (\ref{P})  with the same  cycle structure contribute  in the same way \cite{NDBS}, where all $k$-cycles, i.e., all  cyclic permutations of $k$ photons,  correspond a single parameter $g_k \equiv \mathrm{Tr}(\rho_1^k)$, which can be called the   partial  indistinguishability parameter of   $k$ single photons.

By  employing an  approximation of the Haar distributed matrix elements in the ``collision free" regime by independent complex Gaussian random variables with the probability density $p(U_{kl}) = ({M}/{\pi})\exp\{-M|U_{kl}|^2\}$ \cite{AA}, we obtain   $\langle \widetilde{P}_U(\vec{s}\,|\vec{n}\,)- P_U(\vec{s}\,|\vec{n}\,) \rangle  = 0$ (the average difference between a realistic and  the ideal cases is zero). The variance of this  difference can be used to bound the variational distance $\mathcal{V}$.   Indeed, the variational distance  is bounded by  the 2-norm  as follows  $\mathcal{V}^2 \le \left(\sum_{\vec{m}}1\right)\sum_{\vec{m}}[\widetilde{P}_U(\vec{m}|\vec{n}^{(0)})- P_U(\vec{m}|\vec{n}^{(0)})]^2$. For    the single-photon input, $m_l\le 1$, $|\vec{m}|=N$, we get  $\sum_{\vec{m}}1 \le  M^N/N!$.   Averaging $[\widetilde{P}_U(\vec{m}|\vec{n}^{(0)})- P_U(\vec{m}|\vec{n}^{(0)})]^2$ over the Gaussian approximation  (see  Ref. \cite{NDBS}), we obtain~\footnote{There is a change of notations: the  variational distance is scaled by 2 and our $\mathcal{R}_B(\vec{g}\,)$ is  equal to the  $\mathcal{V}(N,\eta)$ of  Ref. \cite{NDBS}.} 
\be
\langle \mathcal{V}^2\rangle \le \sum_{\vec{c}} \frac{\chi(c_1)\left( 1 - \prod_{k=2}^N g_k^{c_k} \right)^2}{\prod_{k=1}^N k^{c_k}c_k!}\equiv  \mathcal{R}_{B}(\vec{g}\,),
\en{E19}
where the summation runs over the cycle structure   $\vec{c} = (c_1,\ldots, c_N)$ of permutations,  i.e., all $\vec{c}$ satisfying  $\sum_{k=1}^N  k c_k = N$,  and $\chi(n) = \sum_{k=0}^n {n!}/{k!} = \int_1^\infty dz z^n e^{1-z}$.  Here    $ \vec{g} = (g_2, \ldots,g_N)$, the set of partial indistinguishability parameters. By employing Chebyshev's inequality  for the variance $Pr(\mathcal{V}  <  \epsilon) \ge  1 -  {\langle \mathcal{V}^2 \rangle}/{\epsilon^2}$  we get the following  scalability condition on the  photon mode mismatch 
\be
\mathcal{R}_{B}(\vec{g}\,)  \le \epsilon^2\delta,
 \en{E18}
 where $\delta$  excludes a fraction of all networks, as before. For  small mode mismatch    $1-g_k \approx k(1-\langle \mathcal{F}_{ph}\rangle)$ \cite{NDBS}, where     $\langle \mathcal{F}_{ph}\rangle\equiv \overline{|\langle \Phi_1|\Phi_2\rangle|}$ is  the  average  fidelity of the single photons. Here    $\langle\omega|\Phi_i\rangle = \Phi_i(\omega,\tau_i)$ is the  spectral function of the photon from the  $i$th source,  with some fluctuating parameter(s) $\tau_i$ (such as,  for example, the time of arrival), and the overline denotes  the averaging in $\tau_1,\tau_2$. For small mismatch, Eq. (\ref{E18}) becomes \cite{NDBS} 
\be
  (1-\langle \mathcal{F}_{ph}\rangle)^2\left( \frac{N^3}{3} -\frac{N^2}{2} +\frac{7N}{6}-1\right) \le  \epsilon^2\delta.
\en{E20}

The   scalability condition    $1-\langle \mathcal{F}_{ph}\rangle =  O(N^{-{3}/{2}})$ apparently  indicates to  a   non-additive behavior of the photon mode mismatch error   with respect to scaling of the number of photons.  Whereas,  additionally to the above analyzed multi-photon components, photon losses, and detector dark counts, the   noise in the  network realization by  $O(N^2)$ optical elements results in the  scalability condition  on the  element fidelity $1-\mathcal{F}_{el} = O(N^{-2})$ \cite{NoisyBS},  i.e.,  showing  an additive behavior.    

 Our results apply  also   to the BS computer with Gaussian states \cite{GaussBS}, i.e., a BS device where more then $N$ input sources are  randomly heralded for single photons. Dividing  the output probability $P_{out}(\vec{m})$ by   the total number of  $N$  distinct  input modes, we can define an equivalent of $\mathcal{V}$   (\ref{E5}) as a sum over the input and output modes.  Then  the bounds $\mathcal{R}_{A,B}$  apply, since they are obviously  independent  of the indices of    input modes.

\textit{Verification of the experimental BS device  -- } 
Assuming that an operational  device satisfying conditions (\ref{E17}) and (\ref{E20}) is available, how one could verify it? To verify unconditionally  that an experimental device  simulates the BS computer output, the test must be non-polynomial for the classical  computing. Otherwise,  there is a black-box simulator (e.g., a program on a   computer) which would pass  the test.  The variational distance to the ideal BS computer, being the only experimental parameter that quantifies the complexity of an experimental BS device in Theorem 1.3,  can   serve as such an  unconditional test (since the detection is  non-adaptive one would have the necessary data anyway).  The variational distance error can be obtained by comparison with the  classical simulations of the ideal BS computer, feasible for up to  $N\sim 30$ photon sources. Whereas the   variational distance test requires an exponential number of  experimental runs~\footnote{However, a classical computer cannot  use this as an opportunity  to gradually improve its approximation to the probability distribution of the BS computer output, since Theorem 1.3 states that even an   \textit{approximate} simulation is also exponentially hard for a classical device.}, particular tests, i.e.,  against a given distribution,  do not need an exponential number of  runs \cite{notUniform}.   For instance, it was shown \cite{notUniform} that the BS output can be verified against the uniform distribution of Ref. \cite{Gogolin} in a polynomial number of runs by  using as a  witness   the product of squared    row-norms of the network matrix.  Recently, an experimental demonstration was also performed \cite{ExpnU}.

Moreover,  one can devise  tests,  serving  as  evidence of the BS computer operation, which are  based on  some  (independently certified) features of the  experimental setup and also  do not require an exponential number of runs.   For instance,  
 the following unitarity test can be applied to the experimental BS device.  The  linear map $\varphi(U)$  defined in  Eq.~(\ref{E2}) is a  unitary map in the Fock space \cite{A1} and preserves the  group property: $\varphi(U_2U_1) = \varphi(U_2)\varphi(U_1)$.   Therefore, for any input $|\Psi,in\rangle$, the relation  $|\Psi,out\rangle = \varphi(U)|\Psi,in\rangle$  is invertible,   with the inverse map given by  $\varphi(U^\dag)= \left[ \varphi(U)\right]^\dag$. Hence, by placing  high-quality mirrors at the  network output (instead of the detectors) one  redirects the photons to pass through the $U^\dag$-network after the  \mbox{$U$-network,} with their  return to the same input modes. One only  checks the absence of the photons in the inputs $N+1,\ldots, M$. This test   verifies  \textit{in situ} if  the first-order   coherence is still preserved  in the  network output. Conditioned on that the input is certified to satisfy the above derived scalability conditions,   the  unitarity  test for a random network  is a  conditional test of the BS device operation (with a statistical error scaling as inverse square-root in the number of runs), since  the  photons must pass the output state of the $U$-network having a classically hard probability distribution. One can devise other, more sophisticated,  conditional tests of the BS device operation.  For instance,  by using  the $N$-th order generalization of the HOM effect \cite{MPI} one can check that the $N$-th order coherence is preserved,   as is recently proposed in Ref. \cite{ZeroProb}.    This test is also a conditional test,  since it verifies the needed $N$-th order coherence \textit{in situ} but, on the other hand, it  is only    polynomial in $N$ because it simply  checks  for  the zero probability in some of the  output configurations and   is independent of the distribution in all  other  output configurations.

This work was supported by the CNPq of Brazil.  Helpful discussions with M. C. Tichy at the initial stage   and a comment   by P. P. Rohde are acknowledged.  The author is greatly indebted to S. Aaronson for illuminating discussion   and to the anonymous Referees for their remarks, which resulted in a substantial  improvement  of the presentation.


\newpage
\appendix

\section{Details of derivation of the bound in Eqs. (6)-(7) on the variational distance error}

Let us  compute the   average value  of  the    variational distance $\mathcal{V}$, Eq. (5) of the main text.    For $M\gg N^2$ any   $N\times N$-dimensional submatrix  of  a Haar-random $M\times M$-dimensional $U$  is made of the  elements approximated  by the    i.i.d. complex Gaussian random variables  with the probability density $p(U_{kl}) = \frac{M}{\pi}\exp\{-M|U_{kl}|^2\}$ \cite{AA1}. A simple way to obtain  the  average of the  probability $P_U(\vec{s}\,|\vec{n}\,)$ \textit{for arbitrary $\vec{s}$ and $\vec{n}$} is to use a formula for the  matrix permanent employing the Fisher-Yates distribution of the contingency tables $T$: $\mathcal{P}(T|\vec{s},\vec{n}\,)  = \frac{\mu(\vec{s}\,)\mu(\vec{n}\,)}{N_i!\mu(T)}$, where $N_i = |\vec{s}|= |\vec{n}|$,  $T$ is a $M\times M$-dimensional matrix such that $\sum_{l=1}^MT_{kl} = n_k$ and $\sum_{k=1}^M T_{kl} = s_l$ (the contingency table), and  $\mu(T) = \prod_{k,l=1}^M T_{kl}!$.  From Ref.  \cite{AsympProb} we have
 \be
\mathrm{per}(U[\vec{n}\,|\vec{s}\,])  = N_i! \sum_{T}\mathcal{P}(T|\vec{s},\vec{n}\,) \prod_{k,l=1}^MU_{kl}^{T_{kl}}.
\en{E7}
Using the Gaussian approximation we get $\langle \prod_{k,l=1}^M U_{kl}^{T_{kl}} (U_{kl}^{T^\prime_{kl}})^*\rangle  = \delta_{T,T^\prime} \frac{\mu(T)}{M^{N_i}}$ (where $\langle \ldots\rangle$ stands for the average over $U$). From Eq. (4) of the main text  and Eq. (\ref{E7}) we get
\be
\langle  P_U(\vec{s}\,|\vec{n}\,) \rangle = \frac{N_i!}{M^{N_i}}\delta_{|\vec{s}|,N_i}\delta_{|\vec{n}|,N_i},
\en{E8}
 valid for $N_i^2\ll M$.

 In the   ``collision  free" regime,  $M\gg N^2$, due to the boson birthday paradox \cite{AA1},   the probability  of photon bunching at the   network output is bounded, on average in the Haar measure,  by     $1 - (\sum_{|\vec{m}|=N}1)/(\sum_{|\vec{n}|=N}1) =  1- \prod_{k=1}^{N-1}(1-{k}/{M})< N(N-1)/2M$.  Similarly, in our case, the    probability  of a bunched output $\vec{s}$ (i.e., there is  $s_l>1$), $P_B (\vec{s}\,) =\sum_{|\vec{n}|=N_i}P_U(\vec{s}\,|\vec{n}\,) P_I(\vec{n}\,)$,   is bounded by  $\overline{N_i(N_i-1)}/2M$, on average in the Haar measure, where the overline denotes the averaging  with respect to the probability  $P_I(N_i)\equiv \sum_{|\vec{n}|=N_i}P_I(\vec{n}\,)$ of $N_i$  photons in the input.   In the case  of small errors, the average number of photons $\overline{N_i}$ in the input  is   close to $N$, hence,  the  Gaussian approximation can be still used in the calculations below.

By using Eq. (\ref{E8}) for  the  averaging  over $U$  we obtain  
\begin{eqnarray}
\label{E9}
&& \langle \mathcal{V}_1\rangle = 1- \sum_{|\vec{m}|=N} \sum_{\vec{s}} P_D(\vec{m}\,|\vec{s}\,) \sum_{\vec{n}}\langle  P_U(\vec{s}\,|\vec{n}\,) \rangle P_I(\vec{n}\,)
\nonumber\\
\nonumber\\
&& = 1- \sum_{|\vec{m}|=N}\sum_{N_i=0}^\infty\sum_{|\vec{s}|=N_i} P_D(\vec{m}\,|\vec{s}\,)\frac{N_i!}{M^{N_i}}\sum_{|\vec{n}|=N_i}P_I(\vec{n}\,)
\nonumber\\
&& \le 1 - e^{-(M-N)\nu}\left(1 - e^{-\nu}r\right)^N p_1^N\left[1 - \frac{N^2}{2M}\right]\nonumber\\
&& \equiv 1-Q\left[1 - \frac{N^2}{2M}\right].
\end{eqnarray}
We have retained only  the terms with  $N_i=N$ from  the sum over $\vec{s}$   in Eq. (\ref{E9}),  used  that    $\sum_{|\vec{n}|=N}P_I(\vec{n}\,) \ge p_1^N$, and the following inequality
\begin{eqnarray}
\label{E10}
&& \sum_{|\vec{m}|=N}\sum_{|\vec{s}|=N} P_D(\vec{m}\,|\vec{s}\,) \frac{N!}{M^{N}}\nonumber\\
&& \ge \frac{M!}{M^N(M-N)!}e^{-(M-N)\nu}\left(1 - e^{-\nu}r\right)^N\nonumber\\
&& \ge  \left[1 - \frac{N^2}{2M}\right]e^{-(M-N)\nu}\left(1 - e^{-\nu}r\right)^N,
\end{eqnarray}
where the single term   $P_D(\vec{m}\,|\vec{m}\,)$ is retained  to bound from below  the sum over $\vec{s}$ (which is also reasonably close to the whole  sum for small errors),  taken into account that the   number of all outputs $\vec{m}$ is  $\frac{M!}{N!(M-N)!}$, and   the fact that $\frac{M!}{(M-N)!} > M^N\left[1 - \frac{N^2}{2M}\right]$.    On the r.h.s. of Eq. (\ref{E9}) we subtract from 1   the  product of the bound $1 - \frac{N^2}{2M}$ on  the average  probability  of the non-bunched output   and $Q$ --  the   probability  that  $N$ detectors have clicked,  $M-N$ detectors had  zero  dark counts, and that $N$ indistinguishable single photons are at the  network input.   

We bound   $\mathcal{V}_2$  we split it  into three parts by dividing into three parts the summation over the indices $\vec{s}$ and $\vec{n}$  in Eq. (3) of the main text. We will write each respective part of $\mathcal{V}_2$ as   $\mathcal{V}_2[\ldots]$, where  the span of the vector indices $\vec{s}$ and $\vec{n}$ from  the respective partial sum replaces the dots in  the brackets.  Since the absolute value of a sum (in this case the sum over $\vec{s},\vec{n}$ inside the absolute value in each term of  the variational distance $\mathcal{V}_2$ with fixed $\vec{m}$)  is less then the sum of the absolute values, we get
\be
\mathcal{V}_2 \le   \mathcal{V}_2\left[\begin{array}{l}\vec{s}=\vec{m}\\ \vec{n}=\vec{n}^{(0)}\end{array}\right] +  \mathcal{V}_2\left[\begin{array}{l}\vec{s}\ne \vec{m}\\ \vec{n}=\vec{n}^{(0)}\end{array}\right]+ \mathcal{V}_2\left[\begin{array}{l}\quad\mathrm{all} \;\vec{s} \\ \vec{n}\ne \vec{n}^{(0)}\end{array}\right],
\en{E11}
where   the first term on the r.h.s. of Eq. (\ref{E11}) contains $P^{(0)}_{out}(\vec{m}\,)$.   By using Eq. (\ref{E8}) we get (noticing that the first term  on the r.h.s., due to $P^{(0)}_{out}(\vec{m}\,)$, is larger) 
\begin{eqnarray}
\label{E12}
&&\left\langle \mathcal{V}_2\left[\begin{array}{l}\vec{s}=\vec{m}\\ \vec{n}=\vec{n}^{(0)}\end{array}\right] \right\rangle  = \sum_{|\vec{m}|=N}\left\{
\left\langle P_U(\vec{m}\,|\vec{n}^{(0)}) \right\rangle \right.-  \nonumber\\
&& \left. - P_D(\vec{m}\,|\vec{m}\,)\left\langle P_U(\vec{m}\,|\vec{n}^{(0)})\right\rangle p_1^N \right\}\nonumber\\
&&  = \sum_{|\vec{m}|=N}\left[1-P_D(\vec{m}\,|\vec{m}\,)p_1^N\right] \frac{N!}{M^N} \le 1 - Q,
\end{eqnarray}
where we have identified  $Q$ of Eq. (\ref{E9}). Similarly as in  Eqs. (\ref{E9})-(\ref{E10}), we obtain
\begin{eqnarray}
\label{E13}
&& \left\langle \mathcal{V}_2\left[\begin{array}{l}\vec{s}\ne\vec{m}\\ \vec{n}=\vec{n}^{(0)}\end{array}\right] \right\rangle  = 1- \sum_{|\vec{m}|=N}\biggl\{P_D(\vec{m}\,|\vec{m}\,)p_1^N  \nonumber\\
&&\times  \left\langle P_U(\vec{m}\,|\vec{n}^{(0)}) \right\rangle\biggr\} \le 1 - Q\left[1 - \frac{N^2}{2M}\right].
\end{eqnarray}
Finally, using the identities $\sum_{\vec{m}}P_D(\vec{m}\,|\vec{s}\,)=1$ and $\sum_{\vec{s}}P_U(\vec{s}\,|\vec{n}\,)=1$, we obtain for the last term in Eq. (\ref{E11})
\begin{eqnarray}
\label{E14}
&&  \mathcal{V}_2\left[\begin{array}{l}\quad \forall\vec{s}\\ \vec{n}\ne \vec{n}^{(0)}\end{array}\right]   = \sum_{|\vec{m}|=N}\sum_{\vec{s}}\sum_{\vec{n}\ne\vec{n}^{(0)}}\biggl\{P_D(\vec{m}\,|\vec{s}\,)P_U(\vec{s}\,|\vec{n})      \nonumber\\
&&\times    P_I(\vec{n}\,) \biggr\}  \le \sum_{\vec{n}\ne\vec{n}^{(0)}}P_I(\vec{n}\,)  = 1 - p_1^N \equiv Q^\prime. 
\end{eqnarray}

Gathering together the contributions (\ref{E9}) and (\ref{E12})-(\ref{E14}) we obtain   an upper  bound on the Haar-average variational distance in Eq. (5) of the main text   (valid for  $M\gg N^2$)  
\be
\langle \mathcal{V}_1+\mathcal{V}_2 \rangle \le  2\left\{1-Q\left[1 - \frac{N^2}{2M}\right] \right\} + 1-Q + Q^\prime\equiv \mathcal{R}.
\en{E15}

A simple  bound on $\mathcal{R}$  follows from   the inequalities $1 - x^N\le N(1-x)$, and $1-x^Ny^M\le N(1-x) + M(1-y)$,  valid for  $x,y\in[0,1]$ and positive integers $N,M$.  We get $1-Q^\prime \le N(1-p_1)$. Setting $x = (1-e^{-\nu}r)p_1$ and $y = e^{-(1-N/M)\nu}$   we have $1-Q  = 1 - x^Ny^M \le N(1-p_1+r) + (M-N)\nu$,
where we have used that $1-x \le 1-p_1+r$ and $1-y \le (1-N/M)\nu$. Since also $1-(1-a)Q\le 1-Q + a$ for any $0\le a,Q\le 1$ we get 
the resulting bound 
\be
\mathcal{R} \le  \frac{N^2}{M} + 3\left[(M-N)\nu +N r \right]+ 4N(1-p_1).
\en{boundR}


\section{Derivation of Eq. (10) for the output probability for  general multi-photon input}
 Set   $a_i(\omega)$ and $b_i(\omega)$,  $i=1,\ldots,M$ to be  the boson operators for the input and the output modes  of the network with frequency $\omega$. We have   $a^\dag_i(\omega) = \sum_{l=1}^M U_{il}b^\dag_l(\omega)$, with the unitary matrix $U$. If we assume that for an input $\vec{n}$  all    $|\vec{n}| = \sum_{i=1}^Nn_i $  input photons are detected at the output, then the detection  probability  is  given by a POVM consisting of the following   operators  \cite{NDBS1} 
 \begin{eqnarray}
\label{Pi}
&&\Pi(\vec{s}\,) = \frac{1}{\mu(\vec{s}\,)} \left[\prod_{\alpha=1}^{|\vec{n}|}\int\rd\omega_\alpha\right]  \left[ \prod_{\alpha=1}^{ |\vec{n}|}b^\dag_{l_\alpha}(\omega_\alpha)\right]|0\rangle\nonumber\\
&&\times \langle 0|\left[\prod_{\alpha=1}^{|\vec{n}|} b_{l_\alpha}(\omega_\alpha)\right],
\end{eqnarray}
where $(l_1,...,l_N)\equiv \{1,...,1,2,...,2,...,M,...,M\}$,  with index $\l_\alpha=j$ appearing $s_j$ times.   In our case the input density matrix $\rho$, corresponding to the input $\vec{n}$ of $N$ sources, is  given as 
\be
\rho = \rho^{(1)}\otimes\ldots\otimes\rho^{(N)}\otimes|0\rangle\langle0|\otimes\ldots\otimes|0\rangle\langle0|,  
\en{rho}
where the  density matrices of the sources $\rho^{(1)},\ldots,\rho^{(N)}$ are  all diagonal in the Fock basis. In general,  $\rho^{(i)}$ can correspond to a $n_i$-photon input with the photon spectral function having some  fluctuating  parameters (as the time of arrival  or phase, for instance). We can expand each individual density matrix  $\rho^{(i)}$ corresponding to $n_i$ photons  as follows 
\begin{eqnarray}
\label{Spectr}
&&\rho^{(i)} = \sum_k p^{(i)}_k |\Phi^{(i)}_k\rangle \langle \Phi^{(i)}_k|, \quad \sum_kp^{(i)}_k =1, 
\\
&& |\Phi^{(i)}_k\rangle = \int\rd{\omega_1}...\int\rd\omega_{n_i}\Phi^{(i)}_k(\omega_1,\ldots,\omega_{n_i})\prod_{j=1}^{n_i}\frac{a^\dag_i(\omega_j)}{\sqrt{n_i!}}|0\rangle.\nonumber
\end{eqnarray}
Note that, by the permutational symmetry of the creation operators $a^\dag_i(\omega_j)$, the spectral function can be always considered  symmetric in the frequencies. Then it satisfies the usual normalization  condition $\int\rd\omega_1\ldots\int\rd\omega_{n_i}|\Phi^{(i)}_k|^2 =1$. Substituting Eq. (\ref{Spectr}) into Eq. (\ref{rho}), the result into the detection probability $P_U(\vec{s}\,|\vec{n}) = \mathrm{Tr}\{\Pi(\vec{s})\rho\}$, using the expansion for the operators $a^\dag_i(\omega) = \sum_{l=1}^M U_{il}b^\dag_l(\omega)$  and evaluating the  inner products similarly as in Appendix A of Ref. \cite{NDBS1}, we obtain after some algebra
\begin{eqnarray}
&&P_U(\vec{s}\,|\vec{n}\,) = \frac{1}{\mu(\vec{s}\,)\mu(\vec{n}\,)}\sum_{\sigma_1}\sum_{\sigma_2}J(\sigma_1^{-1}\sigma_2)\nonumber\\
&& \times\prod_{\alpha=1}^{|\vec{n}|}U^*_{k_{\sigma_1(\alpha)},l_\alpha}U_{k_{\sigma_2(\alpha)},l_\alpha},
\label{EQ14}
\end{eqnarray}
where  $|\vec{n}|= |\vec{s}|$,  the two sums are over permutations $\sigma_{1,2}$ of $|\vec{n}|$ photon modes 
$(k_1,\ldots,k_{|\vec{n}|})$, with index $k_\alpha=i$ appearing $n_i$ times,  the function on the permutation group $J(\sigma)$ is defined as follows
\be
J(\sigma) =\left[\prod_{\alpha=1}^{|\vec{n}|}\int\rd\omega_\alpha\right] G(\omega_1,...,\omega_{|\vec{n}|}\!\mid\! \omega_{\sigma^{-1}(1)},...,\omega_{\sigma^{-1}(|\vec{n}|)}),
\en{J}
where (setting $S_i \equiv \sum_{j=1}^{i-1}n_j$)
\begin{eqnarray}
\label{G}
&& G(\omega_1,...,\omega_{|\vec{n}|}\!\mid\!\omega^\prime_1,...,\omega^\prime_{|\vec{n}|}) = \prod_{i=1}^N{\Phi^{(i)}}^*(\omega_{S_i+1},...,\omega_{S_i+n_i})\nonumber\\
&&\times \prod_{i=1}^N{\Phi^{(i)}}(\omega^\prime_{S_i+1},...,\omega^\prime_{S_i+n_i}),
\end{eqnarray}
(for simplicity of presentation, we assume that each $\rho^{(i)}$ is pure; the mixed case is obtainable by summation with the product of the  probabilities from Eq. (\ref{Spectr})). 

Observe that by introducing the basis  vectors  $|\omega\rangle$  in the Hilbert   space of the frequency, i.e. $\langle \omega_1 ,\ldots, \omega_{n_i}|\Phi^{(i)}\rangle = \Phi^{(i)}_k(\omega_1,\ldots,\omega_{n_i})$  with $  |\omega_1,...,\omega_{|\vec{n}|}\rangle = |\omega_1\rangle\otimes...\otimes|\omega_{|\vec{n}|}\rangle$, the function $J(\sigma)$ can be also cast as follows
\begin{eqnarray}
\label{J2}
J(\sigma) &=&\left[\prod_{\alpha=1}^{|\vec{n}|}\int\rd\omega_\alpha\right] \langle \omega_{|\vec{n}|},...,\omega_{1}|\mathcal{P}^\dag_\sigma  \nonumber\\
&& \times \rho^{(1)}\otimes\ldots\otimes\rho^{(N)}  |\omega_1,...,\omega_{|\vec{n}|}\rangle\nonumber\\
&& = \mathrm{Tr}\left\{\mathcal{P}^\dag_\sigma  \rho^{(1)}\otimes\ldots\otimes\rho^{(N)}  \right\},
\end{eqnarray}
where $\mathcal{P}_\sigma$  is  the permutation operator acting as
\be
\mathcal{P}_\sigma |\omega_1,...,\omega_{|\vec{n}|}\rangle \equiv |\omega_{\sigma^{-1}(1)},...,\omega_{\sigma^{-1}(|\vec{n}|)}\rangle. 
\en{Psigma}
Note that the trace in Eq. (\ref{J2}) is in the  tensor product of $|\vec{n}|$   Hilbert spaces  associated with the frequencies of  spectral decomposition of the multi-photon states and the permutation operator $\mathcal{P}_\sigma$  acts only on the frequencies, whereas the spatial modes in $\rho^{(1)}\otimes\ldots\otimes\rho^{(N)}$ remain fixed. Finally, substituting  Eq. (\ref{J2}) into Eq. (\ref{EQ14}) and performing the summation inside the trace we obtain the expression in  Eq. (10) of the main text, i.e.
\be
{P}_U(\vec{s}\,|\vec{n}\,) = \frac{1}{\mu(\vec{s}\,) \mu(\vec{n}\,)}\mathrm{Tr} \left\{ \mathcal{U}\rho^{(1)}\otimes\ldots\otimes \rho^{(N)}  \mathcal{U}^\dag\right\},
\en{P1}
by introducing  $\mathcal{U} = \sum_\sigma\left[\prod_{\alpha=1}^NU_{k_{\sigma(\alpha)},l_\alpha}\right]\mathcal{P}^\dag_\sigma$. Observe that the expression in Eq. (\ref{P1}) is obviously  positive (it is a positive definite quadratic form with the  vector $X_\sigma \equiv \prod_{\alpha=1}^NU_{k_{\sigma(\alpha)},l_\alpha}$  indexed by  permutation $\sigma$).  It  reduces to the usual expression of Refs. \cite{AA1,S1,A11}, i.e.,  with the   absolute square of the permanent of a submatrix of $U$ replacing the trace in Eq. (\ref{P1}), in the case of the input states from all the sources being some pure states of just a single multi-photon component with zero mode mismatch between the photons (in this case   the input state is symmetric, i.e. it satisfies $\mathcal{P}_\sigma\left[\rho^{(1)}\otimes\ldots\otimes \rho^{(N)}\right]  = \rho^{(1)}\otimes\ldots\otimes \rho^{(N)}  $).

\end{document}